\documentclass[sigconf]{acmart}
\usepackage{url}
\usepackage{tabularx}
\usepackage{enumitem}
\usepackage[table]{xcolor}

\definecolor{rankA}{RGB}{50,205,50} 
\definecolor{rankB}{RGB}{127,255,0} 
\definecolor{rankC}{RGB}{252,201,190} 
\definecolor{rankD}{RGB}{244,119,110} 
\AtBeginDocument{%
  }

\setcopyright{acmlicensed}
\copyrightyear{2026}
\acmYear{2026}
\acmDOI{XXXXXXX.XXXXXXX}
\acmConference[MODELS 2026]{NIER Track- Proceedings of the ACM/IEEE 29th International Conference on Model Driven Engineering Languages and Systems (MODELS)
}{October 04--09, 2026}{Málaga, Spain}
\acmISBN{978-1-4503-XXXX-X/2018/06}

\begin{document}

\title{A Formalism-Aware Reward Loop for Handwritten UML-to-PlantUML Generation%
}

\author{Mersedeh Sadeghi}
\email{mersedeh.sadeghi@uni-koeln.de}
\affiliation{%
  \institution{University of Cologne}
  \city{Cologne}
  \country{Germany}
}

\author{Simon Scholz}
\email{sschol24@smail.uni-koeln.de}
\affiliation{%
  \institution{University of Cologne}
  \city{Cologne}
  \country{Germany}
}

\author{Adrian Psoch-Bajraktari}
\email{bajraktari@cs.uni-koeln.de}
\affiliation{%
  \institution{University of Cologne}
  \city{Cologne}
  \country{Germany}
}



\begin{abstract}
Handwritten UML sketches are common in early software design, but turning them into structured, analysable modelling artefacts still requires manual reconstruction. Vision-language models can generate PlantUML from diagram images, but prompt-based use treats this as image-to-text generation rather than structured model generation. We investigate formalism-aware rewards: feedback signals derived from analysable model representations rather than surface text. In a worked example, we adapt a vision-language model for handwritten UML-to-PlantUML generation using supervised fine-tuning followed by Group Relative Policy Optimisation. Generated PlantUML is compared against target representations, using XMI for class diagrams and control-flow graphs for activity diagrams. Emerging results show that the adapted model improves compilability and conversion quality over the untuned open model and one proprietary baseline, while remaining competitive with a stronger proprietary baseline on class diagrams. The added benefit of the reward-guided stage remains open on the current held-out set. Error analysis and metric-validity results show that modelling acceptability is only partially captured, motivating rewards and evaluations that combine model analysis with human judgement.
\end{abstract}



\maketitle

\begin{center}
\small
\textit{Accepted for publication in the Proceedings of the ACM/IEEE 29th
International Conference on Model Driven Engineering Languages and Systems
(MODELS 2026). This is the accepted author manuscript and may differ from the
final published version.}
\end{center}

\vspace{1em}

\section{Introduction}
\label{sec:intro}

Software models are often first created informally: on whiteboards, in notebooks, or
as quick sketches during design discussions~\cite{baltes2014sketchesanddiagraminpractice}. 
These sketches are useful for
communication, but remain disconnected from model-driven engineering workflows
until manually reconstructed in a modelling tool. This reconstruction step
is tedious and error-prone, especially when the intended output is not merely a
visual diagram but an analysable model representation~\cite{stoerrle2017, walny2011followthatsketch}. Recent vision-language models
offer an appealing shortcut: given a diagram image, they can be prompted to
generate textual modelling code such as PlantUML. Yet treating sketch-to-model
conversion as generic image-to-text generation misses a central property of the
task: the output is a model artefact whose quality depends on syntactic validity,
structural fidelity, and modelling meaning.

This distinction matters because plausible text is not necessarily a valid or useful
model. A generated PlantUML file may look close to a reference while failing to
compile; it may compile while omitting relationships, attributes, or control-flow
branches; or may express a meaning-preserving alternative differing from the
reference representation. Prompt-only use of general-purpose vision-language models
has little access to modelling-specific constraints during generation, and
standard textual or visual similarity metrics capture only part of the
problem: they do not directly ask whether the generated artefact can be parsed,
analysed, and used as a model.

We therefore explore a different view: model analysis itself can provide feedback for
sketch-to-model generation. We call this idea \emph{formalism-aware rewards}. Rather
than scoring generated PlantUML only as text, we convert it into an analysable
representation and compute feedback in terms of the target modelling formalism. The
same representation-based machinery then serves both as an automatic evaluation
metric and as training feedback (Section~\ref{sec:approach}).

We study this idea on handwritten UML-to-PlantUML generation. Our pipeline first uses
supervised fine-tuning to teach a small open vision-language model, Qwen3.5-4B, the
image-to-PlantUML task, then applies group-relative policy optimisation (GRPO) with
formalism-aware reward components, using a dataset of around 500 handwritten class and
activity diagrams paired with PlantUML~\cite{dataset-anon}. We compare the resulting
model against its untuned base and two proprietary baselines (Gemini~3~Flash and
GPT-4.1~Mini) on a held-out set, and conduct a human ranking study to test whether
automatic model-level scores align with perceived modelling quality.

The goal is not a definitive benchmark but to examine whether formalism-aware model
analysis can become part of the training and evaluation loop. We pursue three
research objectives. 

\begin{itemize}[leftmargin=*, topsep=0pt, itemsep=0pt, parsep=0pt, partopsep=0pt]

  \item \textbf{RO1} is to test whether the full adaptation pipeline can make a
  small open model competitive with proprietary models on handwritten UML-to-PlantUML
  conversion, in both first-attempt compilation and conversion quality.
  \item \textbf{RO2} is to examine how far formalism-aware automatic metrics agree
  with human judgements of model quality, and where they diverge.
  \item \textbf{RO3} is to isolate the reward-guided stage and study when and how
  formalism-aware rewards contribute beyond supervised adaptation.
\end{itemize}

Our emerging results are encouraging but deliberately qualified. The adapted model
moves from last to second overall in both automatic and human evaluation, making a
small open model competitive with substantially larger proprietary systems. At the
same time, an SFT-only ablation shows that, on our small held-out set, the specific
added effect of the reward-guided stage cannot yet be separated from supervised
adaptation: formalism-aware rewards are well-defined model-level signals, but how to
make them reliably improve beyond supervised fine-tuning remains open.

Our contribution is threefold: a formalism-aware reward and evaluation scheme that derives feedback from analysable model representations rather than text; emerging evidence, across automatic metrics, a human study, and error analysis, that the resulting adapted model is competitive on the task; and an ablation that clarifies that the added benefit of the reward-guided stage is not yet established on the held-out set.
\vspace*{-1.25mm}
\section{Related Work and Positioning}
\label{sec:related}

UML diagram recognition has long been studied as a vision and parsing problem. Early
systems relied on deterministic image-processing or trained object-detection pipelines to detect diagram elements and export structured representations such as XMI or
PlantUML~\cite{karasneh-etal1-2013,Axt1786365,razinkas2024sketchesusecasediagram}.
These pipelines work under controlled conditions but are brittle under handwriting
variation, overlapping elements, and noisy layouts. Recent work therefore turns to
multimodal language models prompted directly for UML-to-code conversion. Conrardy et
al.~\cite{conrardy2024imageumlresultsimage} show that proprietary vision-language
models often recover the coarse structure of handwritten UML diagrams yet still
produce syntax errors, omissions, and hallucinated elements as complexity grows.
Similar patterns recur for other PlantUML and Mermaid generation tasks, where models capture high-level structure but struggle with auxiliary constructs and strict compilability~\cite{ranjani2025measuringvisualunderstandingtelecom,guernsey2025harnessing}.

A smaller body of work studies task-specific fine-tuning. Bates et
al.~\cite{BATES2025100660} fine-tune LLaVA variants for UML-to-code generation and
report improved syntax-error rates over untuned baselines, but evaluate mainly with
textual or visual similarity and coarse, categorical error counts. Naboichenko et
al.~\cite{naboichenko2026unlockingumlclassdiagram} fine-tune a vision-language model
for question answering over UML class diagrams, showing that targeted adaptation of a
smaller model can outperform larger general-purpose ones on diagram-specific tasks.
These studies support the value of domain adaptation but do not use the analysable structure of the generated UML artefact as a training signal. Related evaluation work, for
example~\cite{conrardy2024imageumlresultsimage,Bari2024EvaluatingLargeLanguageModelsInExercisesOfUMLClassDiagramModeling},
relies on mistake counts, similarity measures, or manually assigned quality dimensions;
these metrics remain either underspecified, surface-oriented, or hard to scale. We instead
evaluate generated PlantUML through representations derived from the modelling
formalism itself. In doing so we build on model-comparison and semantic-differencing
research in model-driven engineering, where tools such as SiDiff/UMLDiff align and difference models
structurally rather than as text~\cite{xing2005umldiff}. Our reward can be
read as a lightweight, task-specific model-comparison metric turned into a training
and evaluation signal for sketch-to-model generation.

Outside UML, reinforcement learning has been used for structured multimodal
generation: table-to-LaTeX and floorplan-to-JSON systems combine supervised
fine-tuning with GRPO-style optimisation using rewards over syntax, structure, or
spatial overlap~\cite{ling2025table2latexrlhighfidelitylatexcode,liu2026floorplanvlmvisionlanguagemodelfloorplan}.
In those domains the reward stage added value after supervised fine-tuning, whereas
our SFT-only ablation shows that such added value does not arise automatically in UML sketch-to-model generation. To our knowledge, prior sketch-to-model work has not used
formalism-aware model comparison, over UML-specific representations such as XMI and
control-flow graphs, as a reinforcement-learning reward for sketch-to-model generation.
\section{Approach: A Formalism-Aware Reward Loop}
\label{sec:approach}

We treat generated PlantUML as a model artifact, not as text. It can be compiled,
parsed into a structured representation, and compared against a reference model;
prompt-based uses of vision-language models often leave this structure implicit,
evaluating outputs mainly by textual similarity or by whether the generated code
compiles. We compute the training
reward instead from the structure of the target modelling formalism, so that it
measures whether the generated \emph{model} is correct rather than whether its
\emph{text} matches. We call these \emph{formalism-aware rewards}.

Both stages train on a dataset of $\sim$500 handwritten UML diagrams spanning
class and activity diagrams, each paired with human-written PlantUML
code~\cite{cas2uml2026}. Each sample
provides the image, the diagram type, and a reference, which we use as ground
truth, already in analysable form: XMI for class diagrams and a control-flow graph
for activity diagrams. At reward time the policy is prompted to convert the image;
only the generated PlantUML is converted into the matching representation and
compared against the stored reference.

\begin{figure}[t]
  \centering
    \vspace{-2mm}
  \includegraphics[width=.90\columnwidth]{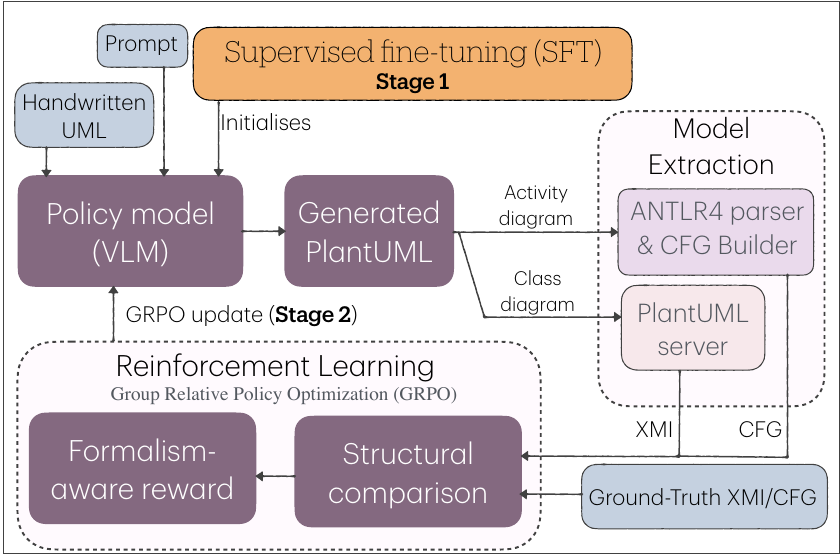}
   \vspace{-2mm}
  \caption{The two-stage, formalism-aware reward loop.}
  \label{fig:reward-loop}
   \vspace{-3mm}
\end{figure}

\paragraph{Two-stage adaptation.} Adaptation proceeds in two stages, shown in Figure~\ref{fig:reward-loop}. The
first stage is supervised fine-tuning, which teaches the model the basic task:
given a handwritten UML image, produce syntactically valid PlantUML that matches
the reference. This stage establishes competence, but it optimises only for
surface agreement with the reference text, not for the structure of the resulting
model.

The second stage addresses exactly that gap. We continue training with
reinforcement learning, using Group Relative Policy Optimization (GRPO), so that
the model is rewarded for the structural correctness of the model it produces. For
each input, the model generates a group of candidate PlantUML outputs. Each
candidate is converted into an analysable representation and scored by the
formalism-aware rewards, giving every candidate a scalar reward. GRPO then
compares the candidates within the group: those scoring above the group average
are reinforced and those below it are discouraged, so the model gradually shifts
probability toward generations whose underlying model structure matches the
reference. Crucially, the learning signal comes not from the generated text but
from the model extracted from it.

To our knowledge, no prior work on UML generation closes the training loop in this
way, deriving the reinforcement signal from analysable representations of the
generated models, such as XMI and control-flow graphs, rather than from textual
similarity.

\paragraph{Reward instantiation.} We instantiate the loop for two diagram types. For class diagrams, PlantUML
exports the generated code to XMI; a generation that fails to compile or yields
unparsable XMI scores zero. Valid outputs are compared with the reference over
classes, attributes, methods, and relationships, so the reward captures structure
that textual similarity cannot, such as whether classes are recovered and whether
associations and inheritance relations connect the correct endpoints.

Activity diagrams have no comparable export, and textual comparison misleads
because one control flow admits many syntactic forms. We parse the generated
diagram into a control-flow graph (CFG), whose nodes are actions and control-flow
constructs and whose edges are execution paths, and compare it with the stored
reference CFG. The reward combines control-flow-graph structure with label
similarity, separating control-flow recovery from wording: a diagram with the
right structure but imperfect labels, or the reverse, still earns partial credit.

Each diagram type's reward sums to 20. A shared format reward (0--1) applies to
both. The class-diagram reward adds a compilation reward (0--1) and structural
rewards over classes (0--3), relationships (0--5), methods (0--5), and attributes
(0--5); the activity-diagram reward adds a compilation reward (0--1), a
control-flow-graph structural reward (0--9), and a label reward (0--9). Because
every component scores a generated model against its reference in the formalism's
own terms, these scores also serve as an automatic, model-level measure of
conversion quality, whose agreement with human judgement we examine in our
evaluation.

\paragraph{Class-diagram reward.} For class diagrams, generated PlantUML is compiled to XMI through the PlantUML
server; if compilation fails or yields no parseable XMI, all content rewards are
zero. Otherwise the class-level reward compares the recovered classes by name and
type, and attribute and method rewards are computed per class by greedily matching
generated members to target members on case-insensitive name, with partial credit for
matching signatures and modifiers. The relationship reward covers six kinds
(association, aggregation, composition, generalisation, realisation, dependency),
matching a generated relationship to a target only when both its kind and its
connected classes agree, with partial credit for multiplicities and role names.

\paragraph{Activity-diagram reward.} The activity-diagram structural reward is computed over the extracted CFG and does not use labels, which are scored separately. It averages two sub-scores: a weighted Jaccard similarity over node-kind multisets, penalising missing or spurious constructs, and an assignment-based topology score inspired by graph edit distance. The latter matches target and generated CFG nodes by minimum-cost bipartite assignment using the Hungarian algorithm; substitution costs distinguish same-kind from different-kind matches, include a swimlane-aware penalty, and compare the outgoing-edge kinds of matched nodes. Label similarity is then computed over matched nodes and edges using TF-IDF cosine similarity rather than exact string matching, so that distinctive domain terms count more than common function words. The artifact provides exact weights, edit costs, and class-diagram sub-scores.

\paragraph{A reusable pattern.} The reward design follows a reusable pattern: parse the model into the formalism's
canonical structure and compare it against the reference. The pattern is general, but
each formalism must be instantiated with its own representation and comparison, as the
class-diagram and activity-diagram rewards above illustrate. We apply it to
class and activity diagrams, two of the most widely used and complementary UML
notations, capturing static structure and dynamic behaviour
respectively~\cite{Koc2021,Reggio2014}; the corresponding analysable
representations are XMI for class diagrams and a CFG for activity diagrams.

\paragraph{Infrastructure.} Reward computation runs inside the training loop, so the reward is available only
after each sampled generation is converted into an analysable representation.
Three components make this practical. A persistent conversion server keeps
repeated PlantUML-to-XMI conversion off the training critical path; a patched
PlantUML processing loop turns malformed generations into deterministic zero
rewards rather than crashes; and a custom ANTLR4 grammar and CFG builder, which we
developed for PlantUML activity diagrams, produce the activity-diagram CFG. These
components are enabling infrastructure; the contribution is the loop that turns
model analysis into a training signal.
\section{Emerging Results}
\label{sec:results}
We evaluate the approach from three complementary angles. The automatic evaluation reuses the formalism-aware reward functions of Section~\ref{sec:approach} as model-level metrics: it measures whether generated PlantUML is syntactically valid and whether its extracted representation preserves the reference structure. The human evaluation assesses perceived modelling quality, which the automatic metrics cannot fully capture. The qualitative error analysis then characterises the mistakes each model makes and explains where automatic and human judgements agree and where they diverge.

All three use a held-out test set of 15 handwritten diagrams per type and compare
four models: our fine-tuned Qwen3.5-4B, the untuned Qwen3.5-4B base model, and two
proprietary baselines, Gemini~3~Flash and GPT-4.1~Mini.
Each model was prompted in
the form it is expected to handle best: the fine-tuned and base models received the
short template used during fine-tuning, while the proprietary baselines received a
longer, more detailed instruction adapted from prior work~\cite{conrardy2024imageumlresultsimage}, even though that work found shorter prompts worked at least as well for their models.
This gives the proprietary baselines a more explicit instruction prompt, making the
comparison conservative with respect to the fine-tuned model.

\subsection{Automatic Evaluation}
\label{subsec:auto}

The automatic evaluation reports the combined \emph{content} score, the sum of the
structural reward components (maximum~19); the format reward
(Section~\ref{sec:approach}) is excluded, as it checks only the response prefix and
not conversion quality. Because the held-out set is intentionally small, we treat
these results as emerging evidence rather than a definitive benchmark comparison.

\paragraph{Compilation.} The most basic requirement is that a generated diagram
compiles at all; non-compiling PlantUML yields no model and is unusable in any
downstream workflow. Adaptation has its largest effect here. Our model compiles
every class diagram (100\%) and raises activity-diagram compilation from the base
model's 20.0\% to 86.7\%, matching Gemini~3~Flash and ahead of GPT-4.1~Mini
(66.7\%); it is the only model to compile all class diagrams
(Table~\ref{tab:results}). Compilation is consistently lower for activity than for
class diagrams across all models, reflecting the tighter syntactic constraints of
control-flow logic.

\paragraph{Conversion quality.} On the combined content score
(Table~\ref{tab:results}), the adapted model moves from last to second on
both diagram types. On class diagrams it reaches 17.60, marginally ahead of
Gemini~3~Flash (17.36) and well above GPT-4.1~Mini (14.97) and the base model
(11.27). On activity diagrams it reaches 13.71, behind Gemini~3~Flash (14.01) but
ahead of GPT-4.1~Mini (9.79) and far above the base model, whose very low score
(2.81) reflects a model barely functional on this type before adaptation. The
absolute gain over the base model is therefore largest on activity diagrams, where
adaptation turns near-failure into usable output.

\paragraph{Component decomposition.} Table~\ref{tab:components} breaks down the automatic score by component. Relative to the untuned base, fine-tuning improves every component; the largest class-diagram gain is in relationship recovery, and the largest activity-diagram gain is compilation, followed by structure and labels. Against the proprietary baselines, no single model dominates: on class diagrams our model leads Gemini~3~Flash and GPT-4.1~Mini on compilation, class-level identification, and attributes, but trails Gemini on relationships and methods; on activity diagrams our model is level with Gemini on compilation but trails on structure and labels, while leading GPT-4.1~Mini throughout. The component scores measure recovery of specific modelling properties, rather than imitation of reference text.Wh

\paragraph{Ablation.} Table~\ref{tab:components} also
lists the supervised-only checkpoint (SFT) as an ablation reference. On the combined
content score, the full SFT-plus-GRPO pipeline is statistically indistinguishable
from supervised fine-tuning alone on both class diagrams ($p=0.94$) and activity
diagrams ($p=0.64$), using paired Wilcoxon tests. The higher SFT figures on the
activity components should therefore not be read as a systematic regression: on this
small held-out set the reward-guided stage demonstrates feasibility but not yet a
statistically detectable gain over supervised fine-tuning.

\begin{table}[t]
  \centering
  \caption{Automatic and human evaluation results.}
  \label{tab:results}
  \footnotesize
  \setlength{\tabcolsep}{3pt}
  \begin{tabular}{@{}lccccccc@{}}
    \toprule
    & \multicolumn{2}{c}{Comp.\ (\%)} & \multicolumn{2}{c}{Content (/19)} & \multicolumn{3}{c}{Borda} \\
    \cmidrule(lr){2-3}\cmidrule(lr){4-5}\cmidrule(lr){6-8}
    Model & Class & Activity & Class & Activity & Overall & Class & Activity \\
    \midrule
    Gemini~3~Flash    & \cellcolor{rankB}93.3  & \cellcolor{rankA}86.7 & \cellcolor{rankB}17.36 & \cellcolor{rankA}14.01 & \cellcolor{rankA}0.759 & \cellcolor{rankA}0.731 & \cellcolor{rankA}0.787 \\
    Ours (fine-tuned) & \cellcolor{rankA}100.0 & \cellcolor{rankA}86.7 & \cellcolor{rankA}17.60 & \cellcolor{rankB}13.71 & \cellcolor{rankB}0.613 & \cellcolor{rankB}0.662 & \cellcolor{rankB}0.564 \\
    GPT-4.1~Mini      & \cellcolor{rankC}86.7  & \cellcolor{rankC}66.7 & \cellcolor{rankC}14.97 & \cellcolor{rankC}9.79  & \cellcolor{rankC}0.442 & \cellcolor{rankC}0.408 & \cellcolor{rankC}0.477 \\
    Base (Qwen)       & \cellcolor{rankD}73.3  & \cellcolor{rankD}20.0 & \cellcolor{rankD}11.27 & \cellcolor{rankD}2.81  & \cellcolor{rankD}0.186 & \cellcolor{rankD}0.200 & \cellcolor{rankD}0.172 \\
    \bottomrule
  \end{tabular}
\end{table}

\begin{table}[t]
  \centering
  \caption{Per-component automatic scores on the held-out set.}
  \label{tab:components}
  \footnotesize
  \setlength{\tabcolsep}{3.5pt}
  \begin{tabular}{@{}l cccc | c@{}}
    \toprule
   Component & Gemini 3 Flash&Ours & GPT-4.1 Mini& Base(Qwen) & SFT \\
     &  &  &  &  & (no RL) \\
    \midrule
    \multicolumn{5}{@{}l}{\textit{Class diagrams}} \\
    Compilation    & \cellcolor{rankB}93.3 & \cellcolor{rankA}100.0 & \cellcolor{rankC}86.7 & \cellcolor{rankD}73.3 & 100.0 \\
    Class-level    & \cellcolor{rankB}92.2 & \cellcolor{rankA}98.3  & \cellcolor{rankC}84.9 & \cellcolor{rankD}70.3 & 98.3  \\
    Relationship   & \cellcolor{rankA}88.9 & \cellcolor{rankB}84.2  & \cellcolor{rankC}64.2 & \cellcolor{rankD}37.6 & 83.6  \\
    Attribute      & \cellcolor{rankB}93.3 & \cellcolor{rankA}98.0  & \cellcolor{rankC}84.0 & \cellcolor{rankD}65.3 & 98.0  \\
    Method         & \cellcolor{rankA}91.0 & \cellcolor{rankB}90.9  & \cellcolor{rankC}82.8 & \cellcolor{rankD}65.6 & 91.2  \\
    \addlinespace
    \multicolumn{5}{@{}l}{\textit{Activity diagrams}} \\
    Compilation    & \cellcolor{rankA}86.7 & \cellcolor{rankA}86.7  & \cellcolor{rankC}66.7 & \cellcolor{rankD}20.0 & 100.0 \\
    Structural     & \cellcolor{rankA}75.1 & \cellcolor{rankB}73.4  & \cellcolor{rankC}50.9 & \cellcolor{rankD}16.0 & 86.0  \\
    Label          & \cellcolor{rankA}70.9 & \cellcolor{rankB}69.2  & \cellcolor{rankC}50.5 & \cellcolor{rankD}13.0 & 78.6  \\
    \bottomrule
  \end{tabular}
    \vspace{-3mm}
\end{table}

\vspace*{-1.25mm}
\subsection{Human Evaluation}
\label{subsec:human}

Automatic rewards measure whether generated PlantUML is valid and structurally close to the reference, but they cannot fully determine whether a generated diagram
is a good interpretation of the sketch. We therefore conducted a human ranking study
over the same four models. The 30 held-out sketches were split into three blocks of
ten, each block balanced with five class and five activity diagrams; each of the 26
participants was randomly assigned one block and ranked the four models' outputs for
its ten sketches, so all 30 sketches were covered across blocks. We capped each
participant at ten sketches because ranking more is cognitively demanding.

Each rater saw the original handwritten sketch alongside the four models' outputs rendered as diagram
images, and ranked them from best to worst; an output that failed to compile was
shown as a rendering-error image rather than a diagram. The reference model was not
shown, so raters judged how well each output captured the sketch rather than how
closely it matched the reference. We aggregate the
rankings with a Borda count, which converts each ranking into points (three for
first place down to zero for last), sums them, and normalises the total to $[0,1]$
by the maximum attainable score; we relate the result to
automatic scores by Spearman correlation. Participants were recruited via SurveyCircle and social media and screened with a
UML-knowledge test, retaining only those above a competence threshold to reach
the final sample of 26. The sample skewed toward computing backgrounds (21 of 26 in
CS/Business Informatics/SE) and early-career education levels (10 current
Bachelor's students, 3 doctoral candidates). Seventeen had hands-on UML experience
and nine had encountered it without using it; none reported no exposure,
consistent with the screening criteria. Inter-rater agreement within groups was
strong (Kendall's $W$ between $0.80$ and $0.91$), which lends weight to the
aggregate ordering.

\paragraph{Ranking.} The human ranking confirms the main trend of the automatic
evaluation: the adapted model moves from last to second overall, ahead of
GPT-4.1~Mini and behind Gemini~3~Flash in perceived quality
(Table~\ref{tab:results}); its mean rank improves from 3.44 (last of four) to 2.16,
behind Gemini's 1.72. This agreement provides evidence that the adapted model's
gains are visible to human raters, not only to the automatic metrics. The human
preference for Gemini is wider on activity diagrams than on
class diagrams, mirroring the automatic scores, where our model is closest to
Gemini on class diagrams and trails further on activity.

\paragraph{Metric validity.} The combined content score
correlates positively and significantly with the human rankings ($\rho = 0.565$,
$p < 0.001$, $n = 120$ model-diagram pairs), and similarly within each diagram type
($\rho = 0.607$ for class, $0.636$ for activity). The association
is \emph{significant but only moderate}: a higher automatic score reliably
coincides with a better human rank, yet the two do not measure quite the same
thing. Two asymmetries make the gap concrete. First, on class diagrams our model and
Gemini~3~Flash score almost the same (17.60 and 17.36) yet human raters still
prefer Gemini, so the metric misses something the raters weigh. Second, compilation success predicts
human preference far more strongly for activity diagrams ($\rho = 0.689$) than for
class diagrams ($\rho = 0.444$): activity diagrams are syntactically harder to
compile, so compiling at all is a stronger signal of quality there, whereas almost
all class diagrams compile and compilation discriminates little.

\subsection{Qualitative Error Analysis}
\label{subsec:errors}

To understand the remaining weaknesses of the adapted model, we compared each
handwritten diagram, its reference model, and the four generated outputs across all
120 outputs (four models on 30 diagrams), grouping recurring mistakes with an
inductively derived taxonomy covering failures of extraction, compilation,
recognition, typing, structure, hallucination, omission, and cosmetic fidelity. The
taxonomy is an organising scheme rather than a measurement instrument: given the
interpretive nature of the labelling and the modest sample, we characterise dominant
patterns rather than report per-category frequencies.

The dominant patterns differ by diagram type. In class diagrams, remaining errors
mostly concern local modelling details, such as missing members, wrong relationship
types or endpoints, or details hard to read in dense sketches. In activity diagrams,
they often concern control-flow interpretation, such as substituting one branching
or looping construct for another or attaching a label to the wrong flow part.
Hallucinations and omissions recur across all models, while the untuned base
produces the most spurious structure.

\begin{figure*}[t]
\centering
\includegraphics[width=\textwidth]{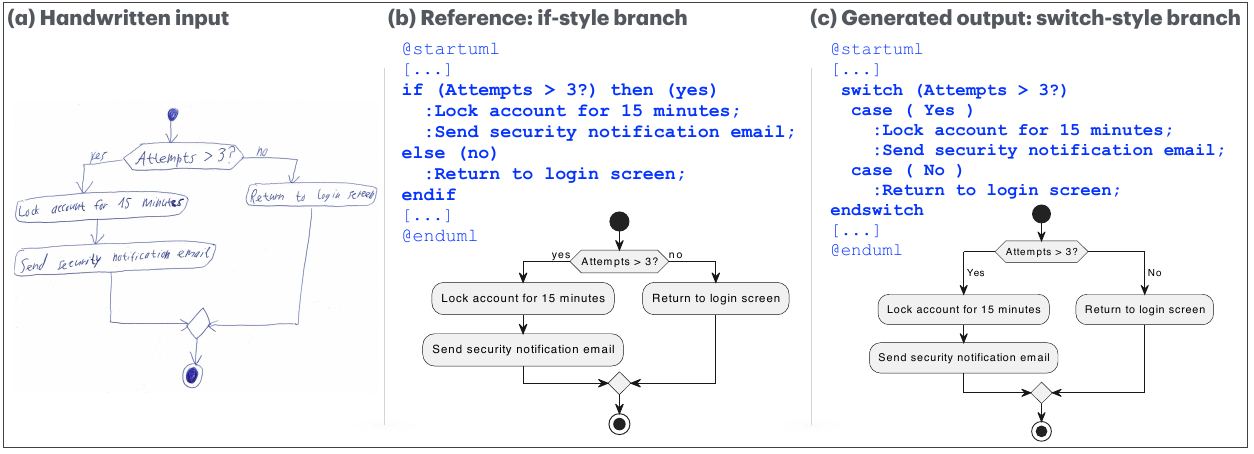}
  \caption{A meaning-preserving deviation: from the handwritten sketch (a), the reference uses an if-branch (b) and the model an equivalent switch-branch (c). The two render to identical diagrams, yet the structural metric penalises the switch.}
  \label{fig:if-switch-meaning-preserving}
\end{figure*}

The main lesson from this analysis is the distinction between
\emph{meaning-preserving} deviations and \emph{meaning-altering} errors. The former changes the representation while keeping the intended interpretation: representing a multi-way branch as a sequence of conditionals, for
example, may preserve the intended control-flow logic even though the parser and the reward functions treat the two forms as different structures. 
Figure~\ref{fig:if-switch-meaning-preserving} shows a representative case: the model renders the sketch's decision as a switch-branch where the reference uses an if-branch. Both produce structurally identical diagrams and preserve the decision logic, yet the control-flow-graph reward scores the switch as a deviation. Such meaning-preserving differences account for part of the gap between the automatic metrics and human raters.

A meaning-altering error, by contrast, changes the logic the diagram represents.
Representation-based rewards penalise both whenever they differ from the target representation, whereas human raters need not: in optional explanations, participants often tolerated meaning-preserving deviations while penalising
meaning-altering ones. 
This is the gap that the moderate metric-human correlation reflects, separating genuine model limitations from the strictness of the chosen representation. Tellingly, meaning-altering errors cluster where the model must read fine visual detail, such as a multiplicity on a dense diagram, pointing to a limit
in visual grounding rather than in PlantUML generation.

\vspace{-2mm}
\subsection{Lessons Learned}
\label{subsec:lessons}

Returning to the objectives of Section~\ref{sec:intro}, three lessons emerge.
On RO1, the full adaptation pipeline yields a competitive model overall, moving the
small open model from last to second among four, ahead of GPT-4.1~Mini and
behind Gemini~3~Flash. On RO3, however, the SFT-only ablation shows that the
specific contribution of the reward-guided stage remains open: on this small
held-out set, supervised fine-tuning and the full pipeline are statistically
indistinguishable. Establishing whether and how model-level rewards help beyond
supervised adaptation is the question taken up in Section~\ref{sec:future}.

A cross-cutting lesson, beyond the three objectives, is that while the idea of
deriving rewards from analysable model representations is generic, the
representation must suit each diagram type. XMI fits class diagrams, whose elements
(classes, attributes, methods, relationships) can be extracted and compared
directly; activity diagrams need a control-flow graph, which compares process
structure better than raw PlantUML text but also exposes the difficulty of telling
an equivalent control-flow formulation from a genuinely different one. Designing
the analysable representation per formalism is therefore part of the method, not a
generic text-matching step, and extending it to further notations is a natural next
step.

On RO2, formalism-aware metrics are scalable but partial proxies for modelling
quality. They capture important downstream properties, especially compilability and
recoverable structure, but they penalise meaning-preserving variation that human
modellers accept and miss the finer cues raters weigh. This argues for evaluation
that pairs formalism-aware analysis with human judgement, and for reward designs
that distinguish meaning-preserving from meaning-altering differences rather than
scoring every structural deviation as an error. Together with the visual-grounding
limitation identified above, these lessons frame the agenda we turn to next.
\section{Future Plans}
\label{sec:future}

Our results suggest that the full adaptation pipeline can yield a small open model
competitive on sketch-to-model conversion, while also exposing two limits: automatic
metrics correlate with human judgement only moderately, and the added effect of the
reward-guided stage remains unresolved on the current held-out set. This points to a
research agenda rather than a single missing feature: making the reward, the
optimisation procedure, the model, and the evaluation reason about meaning rather
than surface structure.

\paragraph{Richer reward optimisation.} One direction is to make the reward more
meaning-aware. A representation-based reward treats any deviation from the stored
target as an error, whereas a modeller accepts semantically equivalent
formulations. This matters most during training: an acceptable meaning-preserving
generation, such as the switch-branch of Figure~\ref{fig:if-switch-meaning-preserving},
can be discouraged by the current reward even though it faithfully captures the
sketch. Future rewards should therefore grant credit for semantic equivalence,
comparing models up to meaning-preserving transformations rather than by structural
identity alone. A second, complementary direction concerns how multiple reward
components are optimised. Because our reward sums several subscores (five for class
diagrams, three for activity diagrams), group-relative optimisation that
normalises the aggregate can let strong components mask weak ones, weakening the
training signal for weaker subscores. Decoupling the normalisation per component,
as in group reward-decoupled policy optimisation (GDPO)~\cite{liu2026gdpo}, may
provide a more targeted signal for the individual constraints our component
analysis shows are recovered unevenly. More generally, reward-guided fine-tuning
should be treated as an optimisation problem in its own right, including component
and diagram-type balancing, regularisation against the supervised policy, and
early stopping.

\vspace{-1.75mm}
\paragraph{Visual grounding.} Our error analysis traced many remaining
meaning-altering errors not to PlantUML generation but to reading fine visual
detail, such as a multiplicity or an arrowhead that distinguishes a generalisation
from a directed association. Addressing this calls for stronger visual grounding:
spatially aware encodings that localise elements within the image, training data
annotated with element positions rather than image-code pairs alone, and agentic
zoom-into-image recognition that inspects dense regions at higher resolution. The
last is a natural fit for the recognition errors we observed, where a method or
attribute name is substituted with a visually similar one. Because this limit is
perceptual, it is unlikely to be solved by reward aggregation alone; gains in visual
grounding should instead compound with better reward design.

\paragraph{Generalisation to further formalisms.} The reward loop is not specific to
the two notations we study. Many modelling languages with analysable
representations could support the same recipe: parse the generated artefact into
that representation and derive feedback from its fidelity to a target
representation. Extending it to further UML notations such as state machines and
sequence diagrams, and to non-UML languages such as BPMN or SysML, would test how
far formalism-aware rewards transfer and what representation each formalism
requires.

\vspace{-1mm}
\paragraph{Towards interactive modelling.} Finally, the divergence between
automatic rewards and human judgement is itself an opportunity. Rather than
treating generation as a single shot evaluated in batch, a system could present its
output, flag elements its own model analysis finds uncertain or inconsistent, and
let the modeller correct them. Those corrections supply exactly the meaning-aware
signal a representation-based reward lacks, and could feed back as preference data
for further training. Combined with a larger and more diverse evaluation — left to
future work given the deliberately small held-out set used here — this would move
formalism-aware models from a one-shot converter towards an interactive modelling
assistant embedded in real workflows.

\section{Conclusion}
\label{sec:conclusion}


This paper argued that sketch-to-model generation should be treated as structured model generation rather than image-to-text generation. We explored formalism-aware rewards as one way to turn model analysis into feedback, using XMI comparison for class diagrams and control-flow-graph comparison for activity diagrams. The adapted open model becomes competitive with larger proprietary baselines, but the added benefit of the reward-guided stage beyond supervised fine-tuning remains unresolved on the current held-out set. More broadly, the results show that model-level metrics
capture important aspects of compilability and structure, while still falling short of human judgement on modelling meaning.

\section*{Data and Artifact Availability}
Code, prompts, reward definitions, and evaluation material are available in an
anonymous repository: \url{https://anonymous.4open.science/r/uml-to-plantuml-F938}.

\bibliographystyle{ACM-Reference-Format}
\bibliography{citations}

@misc{conrardy2024imageumlresultsimage,
      title={From Image to UML: First Results of Image Based UML Diagram Generation Using LLMs}, 
      author={Aaron Conrardy and Jordi Cabot},
      year={2024},
      eprint={2404.11376},
      archivePrefix={arXiv},
      primaryClass={cs.SE},
      url={https://arxiv.org/abs/2404.11376}, 
}

@INPROCEEDINGS{Reggio2014,
  author={Reggio, Gianna and Leotta, Maurizio and Ricca, Filippo and Clerissi, Diego},
  booktitle={2014 2nd International Conference on Model-Driven Engineering and Software Development (MODELSWARD)}, 
  title={What are the used activity diagram constructs? a survey}, 
  year={2014},
  volume={},
  number={},
  pages={87-98},
  doi={}}

@Article{Koc2021,
AUTHOR = {Koç, Hatice and Erdoğan, Ali Mert and Barjakly, Yousef and Peker, Serhat},
TITLE = {UML Diagrams in Software Engineering Research: A Systematic Literature Review},
JOURNAL = {Proceedings},
VOLUME = {74},
YEAR = {2021},
NUMBER = {1},
ARTICLE-NUMBER = {13},
URL = {https://www.mdpi.com/2504-3900/74/1/13},
ISSN = {2504-3900},
DOI = {10.3390/proceedings2021074013}
}

@misc{dataset-anon,

  author       = {Anonymous},

  title        = {Dataset Paper Accepted but Not Yet Publicly Available},

  year         = {2025},

  note         = {Full citation withheld for double-anonymous review and will be added in the camera-ready version.},

  howpublished = {Anonymised reference}

}

@inproceedings{cas2uml2026,
  author    = {Simon Scholz and Mersedeh Sadeghi},
  title     = {{CAS2UML}: A Handwritten Sketch-to-PlantUML Dataset for Class and Activity Diagrams},
  booktitle = {Proceedings of the 41st IEEE/ACM International Conference on Automated Software Engineering (ASE 2026)},
  year      = {2026},
  note       = {Accepted for publication}
}

@article{liu2026gdpo,
  title={Gdpo: Group reward-decoupled normalization policy optimization for multi-reward rl optimization},
  author={Liu, Shih-Yang and Dong, Xin and Lu, Ximing and Diao, Shizhe and Belcak, Peter and Liu, Mingjie and Chen, Min-Hung and Yin, Hongxu and Wang, Yu-Chiang Frank and Cheng, Kwang-Ting and others},
  journal={arXiv preprint arXiv:2601.05242},
  year={2026}
}

@INPROCEEDINGS{karasneh-etal1-2013,
  author={Karasneh, Bilal and Chaudron, Michel R.V.},
  booktitle={2013 5th International Conference on Computer Science and Information Technology}, 
  title={Extracting UML models from images}, 
  year={2013},
  volume={},
  number={},
  pages={169-178},
  doi={10.1109/CSIT.2013.6588776}}

@misc{Axt1786365,
   author = {Axt, Monique},
   institution = {Mid Sweden University, Department of Communication, Quality Management, and Information Systems (2023-)},
   school = {Mid Sweden University, Department of Communication, Quality Management, and Information Systems (2023-)},
   title = {Transformation of sketchy UML Class Diagrams into formalPlantUML models},
   year = {2023}
}

@ARTICLE{razinkas2024sketchesusecasediagram,
  author={Ražinskas, Mantas and Miliūnas, Benas and Jurgelaitis, Mantas and Čeponienė, Lina and Bisikirskienė, Lina},
  journal={IEEE Access}, 
  title={Transforming Sketches of UML Use Case Diagrams to Models}, 
  year={2024},
  volume={12},
  number={},
  pages={185826-185837},
  doi={10.1109/ACCESS.2024.3514455}}

@misc{ranjani2025measuringvisualunderstandingtelecom,
      title={Measuring Visual Understanding in Telecom domain: Performance Metrics for Image-to-UML conversion using VLMs}, 
      author={HG Ranjani and Rutuja Prabhudesai},
      year={2025},
      eprint={2509.11667},
      archivePrefix={arXiv},
      primaryClass={cs.LG},
      url={https://arxiv.org/abs/2509.11667}, 
}

@mastersthesis{guernsey2025harnessing,
  author      = {Guernsey, Grant},
  title       = {Harnessing Large Language Models for Automated Software Diagram Generation},
  school      = {University of Cincinnati},
  year        = {2025},
  type        = {Master's thesis},
  url         = {http://rave.ohiolink.edu/etdc/view?acc_num=ucin1746701542674719},
  note        = {OhioLINK Electronic Theses and Dissertations Center}
}

@article{BATES2025100660,
title = {Unified modeling language code generation from diagram images using multimodal large language models},
journal = {Machine Learning with Applications},
volume = {20},
pages = {100660},
year = {2025},
issn = {2666-8270},
doi = {https://doi.org/10.1016/j.mlwa.2025.100660},
url = {https://www.sciencedirect.com/science/article/pii/S266682702500043X},
author = {Averi Bates and Ryan Vavricka and Shane Carleton and Ruosi Shao and Chongle Pan}
}

@misc{naboichenko2026unlockingumlclassdiagram,
      title={Unlocking UML Class Diagram Understanding in Vision Language Models}, 
      author={Artem Naboichenko and René Peinl},
      year={2026},
      eprint={2605.11634},
      archivePrefix={arXiv},
      primaryClass={cs.CV},
      url={https://arxiv.org/abs/2605.11634}, 
}

@inproceedings{Bari2024EvaluatingLargeLanguageModelsInExercisesOfUMLClassDiagramModeling,
author = {De Bari, Daniele and Garaccione, Giacomo and Coppola, Riccardo and Torchiano, Marco and Ardito, Luca},
title = {Evaluating Large Language Models in Exercises of UML Class Diagram Modeling},
year = {2024},
isbn = {9798400710476},
publisher = {Association for Computing Machinery},
address = {New York, NY, USA},
url = {https://doi.org/10.1145/3674805.3690741},
doi = {10.1145/3674805.3690741},
booktitle = {Proceedings of the 18th ACM/IEEE International Symposium on Empirical Software Engineering and Measurement},
pages = {393–399},
numpages = {7},
location = {Barcelona, Spain},
series = {ESEM '24}
}

@misc{ling2025table2latexrlhighfidelitylatexcode,
      title={Table2LaTeX-RL: High-Fidelity LaTeX Code Generation from Table Images via Reinforced Multimodal Language Models}, 
      author={Jun Ling and Yao Qi and Tao Huang and Shibo Zhou and Yanqin Huang and Jiang Yang and Ziqi Song and Ying Zhou and Yang Yang and Heng Tao Shen and Peng Wang},
      year={2025},
      eprint={2509.17589},
      archivePrefix={arXiv},
      primaryClass={cs.AI},
      url={https://arxiv.org/abs/2509.17589}, 
}

@misc{liu2026floorplanvlmvisionlanguagemodelfloorplan,
      title={FloorplanVLM: A Vision-Language Model for Floorplan Vectorization}, 
      author={Yuanqing Liu and Ziming Yang and Yulong Li and Yue Yang},
      year={2026},
      eprint={2602.06507},
      archivePrefix={arXiv},
      primaryClass={cs.CV},
      url={https://arxiv.org/abs/2602.06507}, 
}

@inproceedings{baltes2014sketchesanddiagraminpractice, series={SIGSOFT/FSE14},
   title={Sketches and diagrams in practice},
   url={http://dx.doi.org/10.1145/2635868.2635891},
   DOI={10.1145/2635868.2635891},
   booktitle={Proceedings of the 22nd ACM SIGSOFT International Symposium on Foundations of Software Engineering},
   publisher={ACM},
   author={Baltes, Sebastian and Diehl, Stephan},
   year={2014},
   month=Nov, pages={530–541},
   collection={SIGSOFT/FSE′14} }

@inproceedings{stoerrle2017,
author = {St\"{o}rrle, Harald},
title = {How are Conceptual Models used in Industrial Software Development? A Descriptive Survey},
year = {2017},
isbn = {9781450348041},
publisher = {Association for Computing Machinery},
address = {New York, NY, USA},
url = {https://doi.org/10.1145/3084226.3084256},
doi = {10.1145/3084226.3084256},
booktitle = {Proceedings of the 21st International Conference on Evaluation and Assessment in Software Engineering},
pages = {160–169},
numpages = {10},
location = {Karlskrona, Sweden},
series = {EASE '17}
}

@INPROCEEDINGS{walny2011followthatsketch,
  author={Walny, Jagoda and Haber, Jonathan and Dörk, Marian and Sillito, Jonathan and Carpendale, Sheelagh},
  booktitle={2011 6th International Workshop on Visualizing Software for Understanding and Analysis (VISSOFT)}, 
  title={Follow that sketch: Lifecycles of diagrams and sketches in software development}, 
  year={2011},
  volume={},
  number={},
  pages={1-8},
  doi={10.1109/VISSOF.2011.6069462}}

@inproceedings{xing2005umldiff,
  title={UMLDiff: an algorithm for object-oriented design differencing},
  author={Xing, Zhenchang and Stroulia, Eleni},
  booktitle={Proceedings of the 20th IEEE/ACM international Conference on Automated software engineering},
  pages={54--65},
  year={2005}
}

\end{document}